\documentclass[aps,prl,reprint,twocolumn,superscriptaddress,showpacs]{revtex4-2}

\usepackage{amsfonts}
\usepackage{amsmath}
\usepackage{graphicx}
\usepackage{bm}
\usepackage{amssymb}
\usepackage{dcolumn}
\usepackage{color}
\usepackage{multirow}
\usepackage{booktabs}
\usepackage[colorlinks,
linkcolor=blue,
anchorcolor=blue,
citecolor=blue,
urlcolor=blue,
]{hyperref}
\begin{document}
	\title{Topological superconductivity and large spin Hall effect in the kagome family Ti$ _{6} $X$ _{4} $ (X = Bi, Sb, Pb, Tl, In)}

	\author{Xin-Wei Yi}
	\affiliation{School of Physical Sciences, University of Chinese Academy of Sciences, Beijing 100049, China}
	
	\author{Zheng-Wei Liao}
	\affiliation{School of Physical Sciences, University of Chinese Academy of Sciences, Beijing 100049, China}
	
	\author{Jing-Yang You}
	\email{phyjyy@nus.edu.sg}
	\affiliation{Department of Physics, Faculty of Science, National University of Singapore, 117551, Singapore}

	\author{Bo Gu}
	\email{gubo@ucas.ac.cn}
	\affiliation{Kavli Institute for Theoretical Sciences, and CAS Center for Excellence in Topological Quantum Computation, University of Chinese Academy of Sciences, Beijing 100190, China}

	\author{Gang Su}
	\email{gsu@ucas.ac.cn}
	\affiliation{Kavli Institute for Theoretical Sciences, and CAS Center for Excellence in Topological Quantum Computation, University of Chinese Academy of Sciences, Beijing 100190, China}
	\affiliation{School of Physical Sciences, University of Chinese Academy of Sciences, Beijing 100049, China}

	\begin{abstract}
Topological superconductors (TSC) become a focus of research due to the accompanying Majorana fermions. However, the experimentally reported TSC are extremely rare. The recent experiments reported the kagome TSC AV$_{3}$Sb$_{5}$ (A=K, Rb, Cs), which exhibit unique superconductivity, topological surface states (TSS), and Majorana bound states. More recently, the first titanium-based kagome superconductor CsTi$ _{3} $Bi$ _{5} $ with nontrivial topology was successfully synthesized as a perspective TSC. Given that Cs contributes little to the electronic structures of CsTi$ _{3} $Bi$ _{5} $ and binary compounds may be easier to be synthesized, here, by density functional theory calculations, we predict five stable non-magnetic kagome compounds Ti$ _{6} $X$ _{4} $ (X = Bi, Sb, Pb, Tl, In) which exhibit superconductivity with critical temperature Tc = 3.8$ - $5.1 K, nontrivial $\mathbb{Z}$$_2$ band topology, and TSS close to the Fermi level. In addition, the large intrinsic spin Hall effect is obtained in Ti$ _{6} $X$ _{4} $, which is caused by gapped Dirac nodes and nodal lines due to a strong spin-orbit coupling. This work offers new platforms for TSC and spintronic devices.
	\end{abstract}
	\maketitle
\section{I. INTRODUCTION}
Topological superconductor (TSC) with Majorana fermions becomes an important subject in condensed matter physics \cite{RN529}. Their topological gapless excitations of linear dispersion and particle-hole symmetry can naturally meet the two requirements of Majorana fermion, where excitations obey the Dirac equation and particles are equal to their own anti-particles \cite{RN599}. The Majorana fermion in TSC following non-Abelian statistics is promising for topological quantum computations without decoherence \cite{RN603, RN516}. Two possible approaches to achieve TSC are odd-parity superconductors with inherently strong topology \cite{RN538, RN539} and superconducting proximity effect \cite{ RN440, RN574, RN575}. Experimental detection of surface Andreev bound states and theoretical analysis have given supporting information of the topological superconductivity in the odd-parity superconductors Cu$ _{x} $Bi$ _{2} $Se$ _{3} $ \cite{RN532, RN441, RN535} and Sn$ _{1-x} $In$ _{x} $Te \cite{RN584}. The proximity effect has focused on proximity-induced coupling of s-wave superconductors with topological insulators \cite{RN440} or semiconductors with strong spin-orbit coupling (SOC) \cite{RN574, RN575}. Several artificially fabricated heterostructures and nanowires have revealed evidences of the Majorana fermions along this route, including epitaxial Bi$ _{2} $Te$ _{3} $ films grown on NbSe$ _{2} $, \cite{RN572, RN606, RN573}, InAs nanowires segment with epitaxial Al \cite{RN576}, InSb nanowires contacted with NbTiN \cite{RN577}, Fe atomic chains on the surface of Pb \cite{RN578}. Additionally, some materials with both bulk superconductivity and surface topological Dirac cones can also intrinsically establish this proximity effect \cite{RN375, RN546, RN514, RN601, RN602, RN504}. Majorana zero modes has been observed in this kind of intrinsic TSC, including the iron-based superconductors $ - $ FeTe$ _{1–x} $Se$ _{x} $ \cite{RN533, RN592, RN594, RN536, RN568}, CaKFe$ _{4} $As$ _{4} $ \cite{ RN548}, LiFeAs \cite{RN549,RN604} and van der Waals material 2M-WS$_2$ \cite{RN600}. In terms of fabrication, the intrinsic TSC are more promising than the heterostructures and nanowires. Additionally, both experimental and predicted intrinsic TSC are extremely rare and most of them can only achieve superconductivity or suitable topological surface states (TSS) near the Fermi energy (E$ _{F} $) by doping. Finding more intrinsic TSC candidates with high Tc and TSS in the vicinity of E$ _{F} $ is highly urgent.

Recently, robust Majorana bound states were observed in a new kagome superconductor CsV$ _{3} $Sb$ _{5} $ \cite{RN1} as a new paradigm of intrinsic TSC. This new supercondctor family AV$ _{3} $Sb$ _{5} $ (A=K, Rb, and Cs) exhibit unique superconductivity with superconducting transition temperatures (Tc) of 0.9$-$2.5K at ambient pressure \cite{ RN8, RN11, RN64, RN83, RN207, RN202, RN9, RN132, RN52, RN209, RN83, RN54, RN190, RN469}, a nontrivial $\mathbb{Z}$$_2$ topology \cite{RN461, RN8}, and other novel quantum properties \cite{RN5, RN34, RN187, RN451, RN34, RN41, RN120}. Very recently, the first Ti-based kagome CsTi$ _{3} $Bi$ _{5} $ with the AV$ _{3} $Sb$ _{5} $ prototype structure has also been synthesized experimentally \cite{RN565} following theoretical prediction \cite{RN151}, and it exhibits a Tc of about 4.8K, which is much higher than that of AV$ _{3} $Sb$ _{5} $ \cite{RN565}. It was predicted that CsTi$ _{3} $Bi$ _{5} $ has nontrivial band topology and robust TSS, implying a possible TSC similar to AV$ _{3} $Sb$ _{5} $. It is very intriguing to find more candidates of TSC with high Tc and intrinsic TSS in similar Ti-based kagome systems. Considering that the density of states (DOS) in CsTi$ _{3} $Bi$ _{5} $ near E$ _{F} $ is mainly contributed by Ti atoms while Cs has almost no contribution and in general binary compounds are also easier to be synthesized in experiments, it will be interesting and imperative to study the Ti-based binary compounds.

In this paper, by density functional theory calculations we predicted kagome non-magnetic family Ti$ _{6} $X$ _{4} $ (X = Bi, Sb, Pb, Tl, In). Ti$ _{6} $X$ _{4} $ are stacked by Ti-based kagome layers and X-based honeycomb layers similar to CsTi$ _{3} $Bi$ _{5} $. The low E$ _{hull} $ in energy convex hull and phonon spectra without imaginary frequency of Ti$ _{6} $X$ _{4} $ show the evidences of their thermodynamic and dynamic stability. The nontrivial $\mathbb{Z}$$_2$ index and corresponding TSS near E$ _{F} $ with spin-moment locked spin textures demonstrate that they are ideal $\mathbb{Z}$$_2$ topological metals. On the other hand, the calculated electron-phonon coupling (EPC) based on the Bardeen-Cooper-Schrieffer (BCS) theory suggests that they have superconducting transitions with a transition temperature Tc of 3.9$ - $5.1K. The coexistence of superconductivity with high Tc and ideal TSS offers promising platforms for realizing TSC and Majorana fermions. Moreover, the band structures of Ti$ _{6} $X$ _{4}$ show abundant Dirac nodes and Dirac nodal lines (DNLs), all of which have gaps in the presence of the strong SOC. The calculated intrinsic spin Hall conductivity (SHC) shows that these gapped nodes and DNLs contribute to a large SHC in Ti$ _{6} $X$ _{4} $, where SHC of Ti$ _{6} $Bi$ _{4}$ and Ti$ _{6} $Sb$ _{4} $ can reach 354 and 629 $ \hbar $·(e·$ \Omega $·cm)$ ^{-1} $, respectively.

\section{II. Results \lowercase{for} T\lowercase{i}$ _{6} $B\lowercase{i}$ _{4}$}

\subsection{A. Crystal structure of Ti$ _{6} $Bi$ _{4} $}
\begin{figure}[t]
	\centering
	\includegraphics[scale=0.345,angle=0]{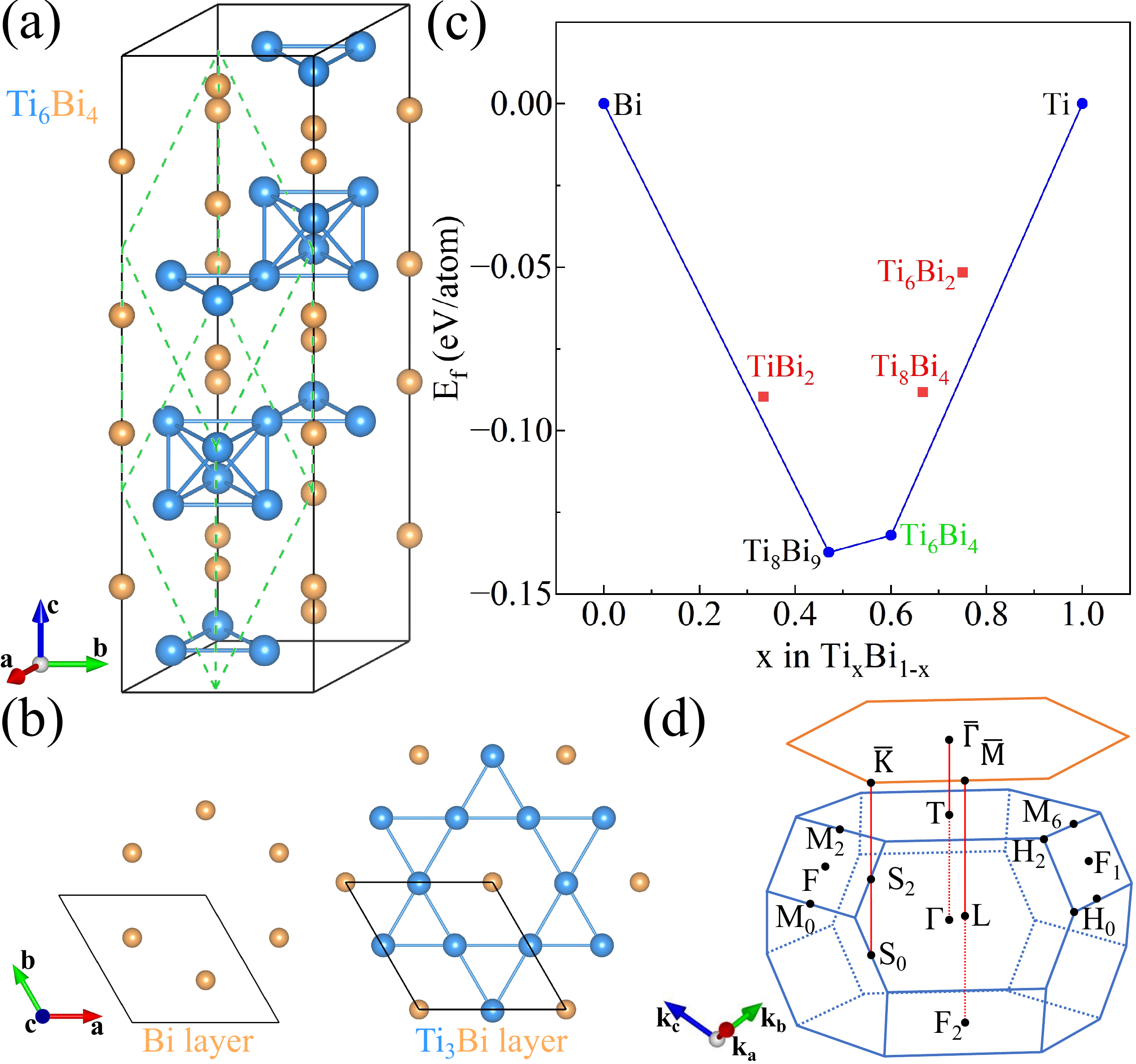}\\
	\caption{(a) The crystal structure for Ti$ _{6} $Bi$ _{4} $ stacked by Ti$ _{3} $Bi and Bi layers. The black solid and green dotted lines represent the Bravais and primitive lattices, respectively. (b) Top view of Bi and Ti$ _{3} $Bi atomic layers. (c) The convex hull of Ti$ _{x} $Bi$ _{1-x} $. (d) BZ with high symmetry points. 3D and 2D BZ are drawn with blue and orange solid lines, respectively.}\label{fig1}
\end{figure}

CsTi$ _{3} $Bi$ _{5} $ is a layered material \cite{RN565}, consisting of Bi and Ti$ _{3} $Bi layers, as shown in FIG.~\ref{fig1}(b). The structure of Ti$ _{6} $Bi$ _{4} $ can be obtained by stacking these atomic layers as seen in FIG.~\ref{fig1}(a), where different Ti$ _{3} $Bi layers stacked in the 'c' direction have in-plane sliding and the two Ti$ _{3} $Bi layers are sandwiched by Bi honeycomb layers. It should be noticed that the Ti kagome nets have some breathing distortions, where the triangles of kagome nets are divided into two unequal sizes. Ti$ _{6} $Bi$ _{4} $ has a rhombohedral structure with a space group of R$ \bar{3} $m (No.166). Its Bravais and primitive lattices are represented by black solid and green dotted lines in FIG.~\ref{fig1}(a), respectively. The bulk Brillouin zone (BZ) and high symmetry points are plotted in FIG.~\ref{fig1}(d), where four inequivalent time-reversal invariant momenta (TRIM) points are labeled as $ \Gamma $, T, F and L.

By calculating the formation energies E$ _{f} $ of all Ti-Bi binary systems \cite{RN475, RN501, RN885}, we can draw the convex hull as shown in FIG.~\ref{fig1}(c). We can see that Ti$ _{6} $Bi$ _{4} $ is thermodynamically stable with E$ _{hull} $ = 0meV. Among all these binary Ti-Bi compounds listed in Fig. 1(c), only Ti$ _{6} $Bi$ _{4} $ has kagome nets. Its special structure similar to CsV$ _{3} $Sb$ _{5} $ and CsTi$ _{3} $Bi$ _{5} $ makes Ti$ _{6} $Bi$ _{4} $ become promising materials to explore the possible intrinsic topological superconductors. The experimental synthesis of a new phase of Ti$ _{3} $Bi$ _{2} $ was reported, where the elements ratio of Ti:Bi = 3:2 was obtained by the energy dispersive x-ray spectroscopy but its exact structure has not been determined \cite{RN525}. It indicates that our predicted Ti$ _{6} $Bi$ _{4} $ may already be fabricated in the experiment.

\subsection{B. Electronic band structure and topological properties}

\begin{figure}[t]
	\centering
	\includegraphics[scale=0.47,angle=0]{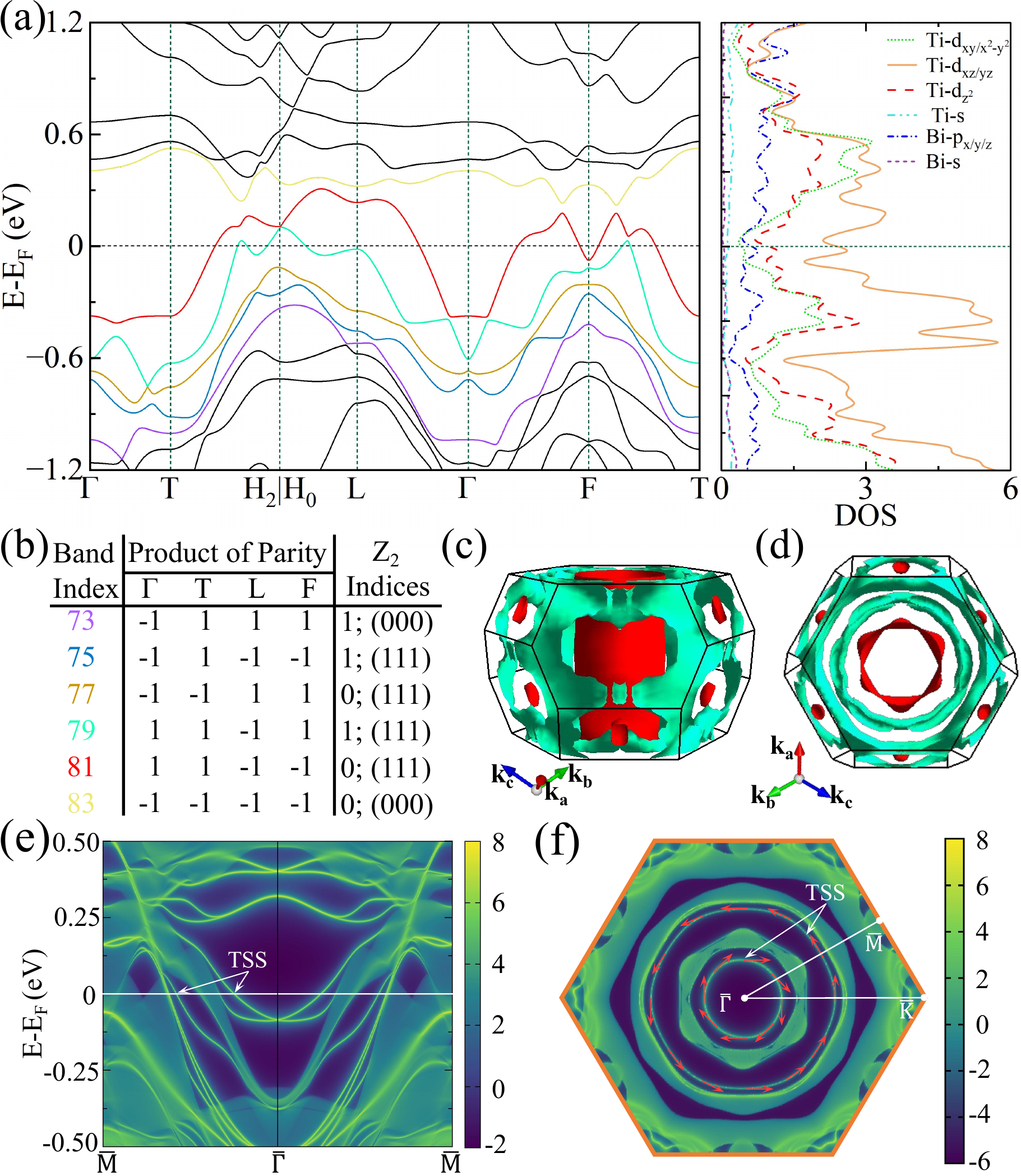}\\
	\caption{(a) The electronic band structure and partial DOS with SOC for Ti$ _{6} $Bi$ _{4} $. Different bands near the E$ _{F} $ are drawn in different colors. (b) Product of parity for four inequivalent TRIM points and $\mathbb{Z}$$_2$ index of bands near Fermi level. Three-dimensional Fermi surface of Ti$ _{6} $Bi$ _{4} $ in (c) side and (d) top views. (e) The surface spectrum functions along $ \bar{M} $-$ \bar{\Gamma} $-$ \bar{M} $ paths projected on (001) plane for Ti$ _{6} $Bi$ _{4} $. (f) Spin texture of TSS projected on (001) plane for Ti$ _{6} $Bi$ _{4} $ at E-E$ _{F} $ = 0meV.}\label{fig2}
\end{figure}

The electronic energy bands and partial DOS with SOC for Ti$ _{6} $Bi$ _{4} $ are plotted in FIG.~\ref{fig2}(a). Comparing the electronic band structure without SOC in Fig. S1, SOC lifts the degeneracy at the Dirac nodes and DNLs, which generates a continuous band gap between two adjacent energy bands in the whole BZ. The DOS exhibits an obvious valley near the Fermi energy and the projected DOS of Ti atoms is much larger than that of Bi atoms. The band structure with weights of projected different orbitals of Ti and Bi atoms is shown in Fig. S2 in Supplemental Information. The bands near the Fermi level are dominated by the five d orbitals of Ti and the three p orbitals of Bi, while the s orbitals of Ti and Bi have little contributions. Experimental angle-resolved photoemission spectroscopy (ARPES) measurement and the density functional theory (DFT) calculations of the V-based kagome structures quantitatively give similar electronic band structures, indicating the validity of the calculated band structures by DFT in these types of systems \cite{RN8}. 
 
With the time-reversal and inversion symmetries of Ti$ _{6} $Bi$ _{4} $, its strong $\mathbb{Z}$$_2$ topological invariant can be calculated by the parity of wavefunctions at all TRIM points \cite{RN372}. Moreover, other three weak $\mathbb{Z}$$_2$ topological invariants can also be calculated. It can be seen from FIG.~\ref{fig2}(b) that several energy bands, including bands 73, 75 and 79 below the Fermi energy (E$ _{F} $) have a strong $\mathbb{Z}$$_2$ index, resulting in abundant clear Dirac cone TSS near the Fermi level in the surface spectrum functions as shown in FIG.~\ref{fig2}(e). The nontrivial TSS and nontrivial $\mathbb{Z}$$_2$ index make Ti$ _{6} $Bi$ _{4} $ a $\mathbb{Z}$$_2$ topological metal.

Unlike most VSb-based kagome structures where the TSS are usually submerged in their bulk states, the bulk states of Ti$ _{6} $Bi$ _{4} $ have a large band gap at ‘$ \Gamma $’ point near E$ _{F} $, which makes its TSS clearly exist without entanglement with bulk states as shown in FIG.~\ref{fig2}(a). The 3D Fermi surface (FS) is plotted in FIGs.~\ref{fig2}(c, d), where the electron pockets are concentrated near F, F$ _{1} $, F$ _{2} $, $ \Gamma $ and T points. The electron pocket centered on the $ \Gamma $ point shows an obvious cylindrical surface along the direction perpendicular to the kagome net, which enables ARPES to easily measure the TSS near $ \bar{\Gamma} $. With calculated surface Green’s function for a semi-infinite system and the surface spectrum function, the spin texture of surface states at fixed energy can be directly obtained. The detailed methods can be seen in Supplemental Information. Therefore, we further draw the projected surface spectral functions and spin textures on (001) plane at E$ _{F} $ in FIG.~\ref{fig2}(f). The TSS form multiple circles centered on the $ \bar{\Gamma} $ point, which presents obvious spin-momentum locking, showing the existence of robust TSS near E$ _{F} $ again.

\subsection{C. Superconductivity}
\begin{figure}[t]
	\centering
	\includegraphics[scale=0.26,angle=0]{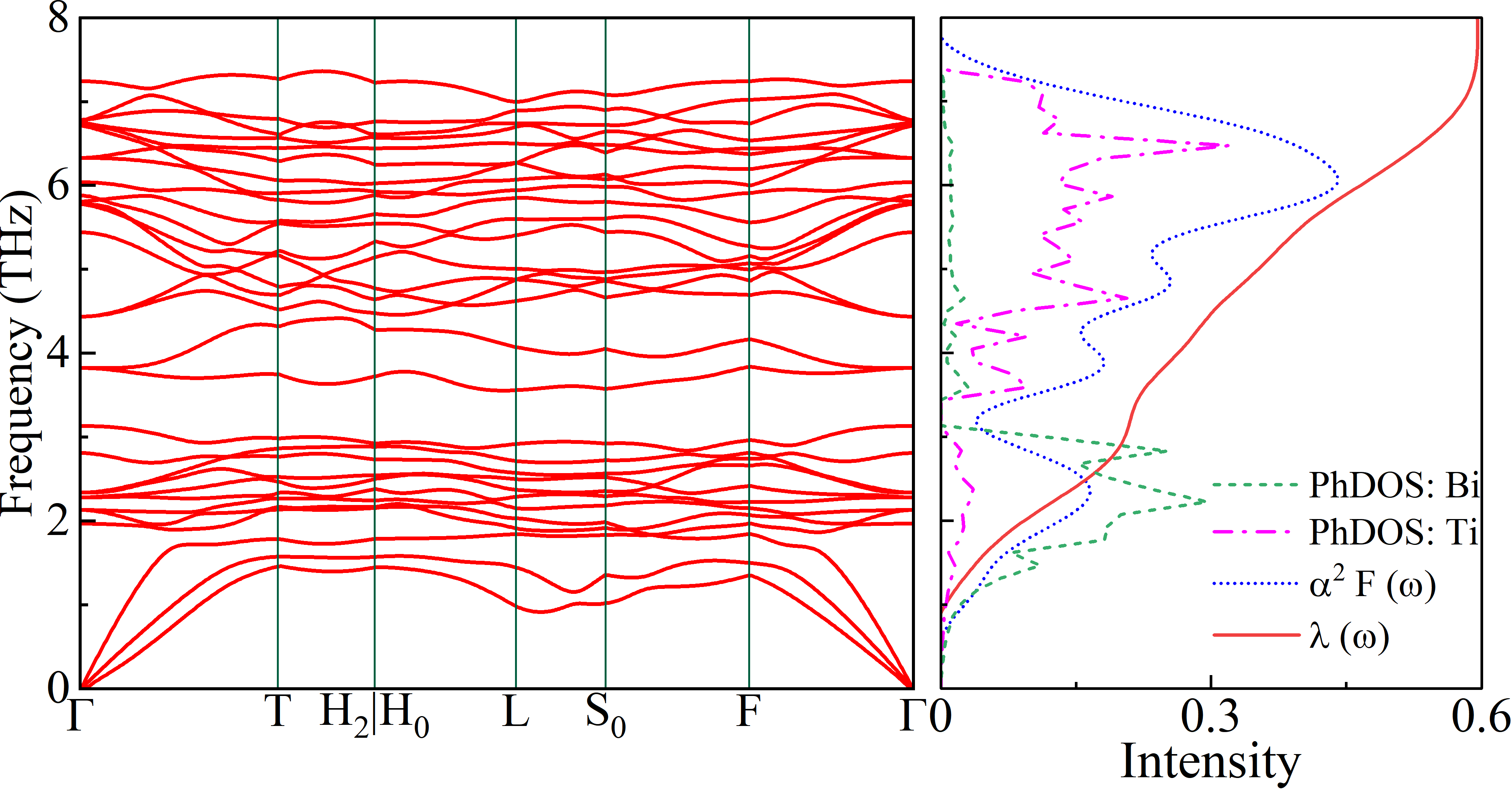}\\
	\caption{The phonon spectrum, projected PhDOS, Eliashberg spectral function $\alpha$$^2$F($\omega$), and cumulative frequency dependent EPC $\lambda$($\omega$) for Ti$_6$Bi$_4$.}\label{fig4}
\end{figure}

Similar to CsV$ _{3} $Sb$ _{5} $ and CsTi$ _{3} $Bi$ _{5} $, the emergence of superconducting ground states in Ti$ _{6} $Bi$ _{4} $ is very promising. To study the superconductivity in Ti$ _{6} $Bi$ _{4} $, we first calculate its magnetic properties. We consider several typical collinear and noncollinear magnetic configurations as shown in FIG. S3. By comparing the total energies and final magnetic moment per atom of these magnetic configurations, Ti$_6$Bi$_4$ can be identified as a nonmagnet. Furthermore, we calculate the Eliashberg spectral function $\alpha$$^2$F($\omega$) and EPC $ \lambda$($\omega$) (see FIG.~\ref{fig4}), then the EPC $ \lambda$($\omega$ = $\infty$) and Tc of Ti$ _{6} $Bi$ _{4} $ are estimated to be 0.586 and 3.8K, respectively, which are relatively high values among the recently discovered kagome superconductors. From FIG.~\ref{fig4}, we find that the phonon DOS (phDOS) at high frequency ($ \sim{6} $THz) and low frequency ($ \sim{2} $THz) is dominated by the contribution of Ti and Bi atoms, respectively. As a rough estimation, the contribution of Ti atoms vibration accounts for more than half of the total EPC. Since the mass of Ti atom is smaller than that of V atom, this partially explains the enhanced EPC and Tc in Ti-based kagome superconductors.  

\subsection{D. Spin Hall effect}

\begin{figure}[t]
	\centering
	\includegraphics[scale=0.53,angle=0]{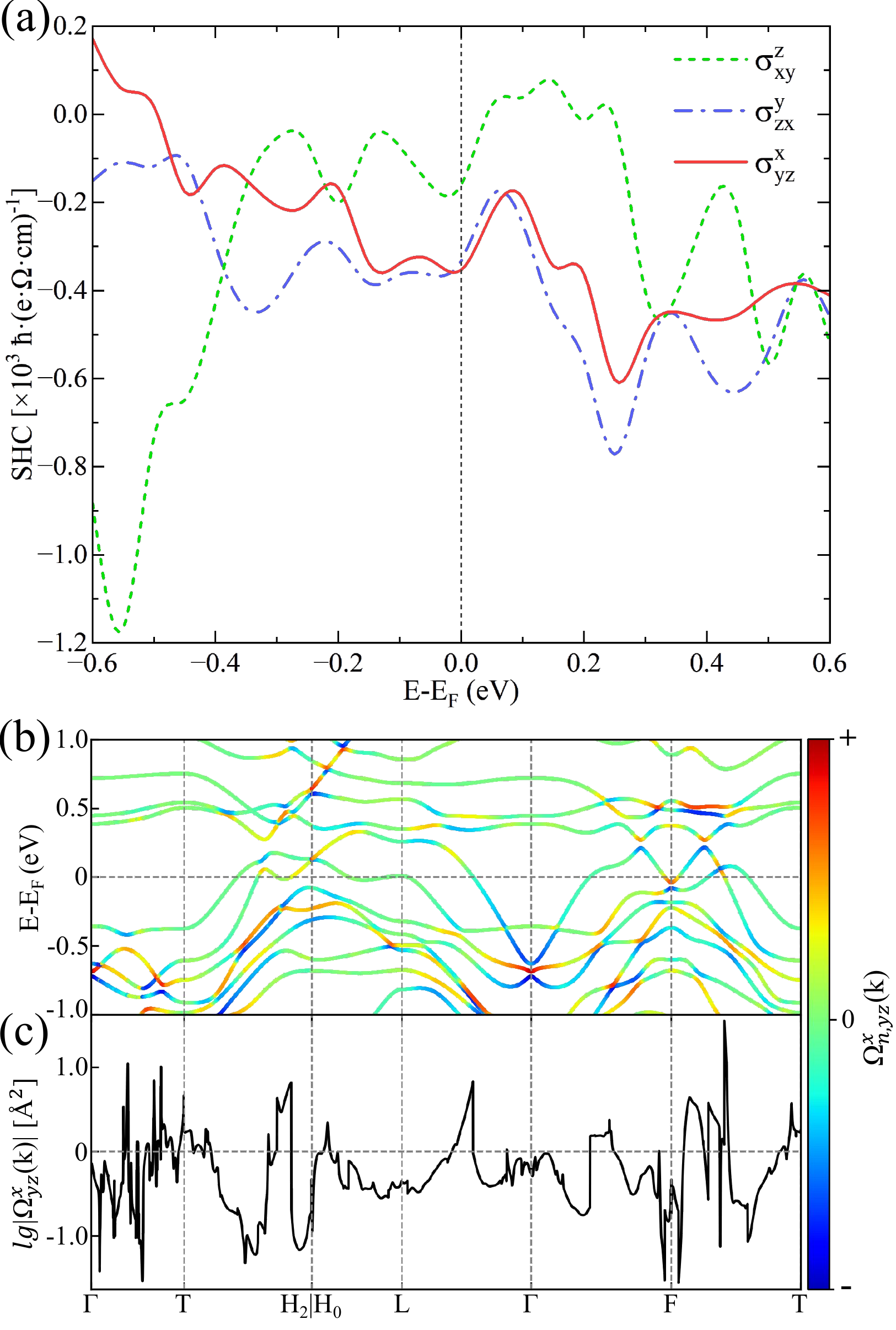}\\
	\caption{(a) Three independent components of SHC tensor as a function of chemical potential for Ti$ _{6} $Bi$ _{4} $. (b) The band structure of Ti$ _{6} $Bi$ _{4} $ weighted by $ \Omega_{n,yz}^{x}(\mathbf{k}) $ and (c) the k-resolved $ \Omega_{ yz}^{x}(\mathbf{k}) $ along the high-symmetry paths at E$ _{F} $.}\label{fig5}
\end{figure}

The symmetry protected DNL in the energy band can serve as a source of various quantum phenomena, such as the spin Hall effect (SHE). The large SOC mixing different spin components of wave functions produces a big numerator, and the small gap of DNL induced by SOC contributes a small denominator in Eq. (4) in Supplemental Information, so a large SHC may appear. Due to the large SOC and corresponding gapped DNLs, a large SHC in Ti$ _{6} $Bi$ _{4} $ is expected. The crystal symmetry of Ti$ _{6} $Bi$ _{4} $ constraints three independent components of SHC tensor, i.e. $ \sigma_{xy}^{z} $ = $ -\sigma_{yx}^{z} $; $\sigma_{zx}^{y} $ = $-\sigma_{zy}^{x} $; $ \sigma_{yz}^{x} $ = $ -\sigma_{xz}^{y} $ \cite{RN490}. Therefore, we evaluate three independent components $ \sigma_{xy}^{z} $, $ \sigma_{zx}^{y} $, and $ \sigma_{yz}^{x} $ as a function of chemical potential as plotted in FIG.~\ref{fig5}(a). It can be seen that the magnitudes of three components are around 160$ - $354 $ \hbar $·(e·$ \Omega $·cm)$ ^{-1} $ at E$ _{F} $. The SHC changes drastically with the change of chemical potential, and $\sigma_{xy}^{z}$ can reach 1168 $ \hbar $·(e·$ \Omega $·cm)$ ^{-1} $ at -0.55eV. 

To reveal the origin of large SHC, we plot the band structure of Ti$ _{6} $Bi$ _{4} $ colored by the magnitude of spin Berry curvature $ \Omega_{n,yz}^{x}(\mathbf{k}) $, as well as the k-resolved $ \Omega_{yz}^{x}(\mathbf{k}) $ at E$ _{F} $ in FIGs.~\ref{fig5}(b, c). It is noted that $ \Omega_{n,yz}^{x}(\mathbf{k}) $ strongly depends on wave vector $ \mathbf{k} $, and $ \Omega_{n,yz}^{x}(\mathbf{k}) $ shows prominent peaks at positions of gapped nodes and DNLs, which mainly contribute to SHC. The reason for relatively isotropic SHC is that Dirac nodes and DNLs are located at different energy levels. The large SHC of Ti$ _{6} $Bi$ _{4} $ provides promising applications for spintronics.

\section{III. Results \lowercase{for} kagome family T\lowercase{i}$ _{6} $X$ _{4} $ (X = S\lowercase{b}, P\lowercase{b}, T\lowercase{l}, I\lowercase{n})}
 To our knowledge, CsTi$ _{3} $Bi$ _{5} $ exhibits many interesting properties and is the only superconductor with titanium-based kagome net \cite{RN565}. Therefore, it is important to find more titanium-based kagome superconductors. In the Ti$ _{6} $Bi$ _{4} $, the electronic states near the E$ _{F} $ are mainly contributed by Ti atoms, which indicates that the substitution of Bi atoms may produce the family Ti$ _{6} $X$ _{4} $, similar to Ti$ _{6} $Bi$ _{4} $. We substitute Bi in Ti$ _{6} $Bi$ _{4}$ with all elements of groups IIIA, IVA, VA and VIA except radioactive polonium, and find four dynamically stable compounds Ti$ _{6} $X$ _{4} $ (X = Sb, Pb, Tl, In). Furthermore, DFT calculations show that Ti$ _{6} $Sb$ _{4} $, Ti$ _{6} $Pb$ _{4} $, and Ti$ _{6} $Tl$ _{4} $ are all thermodynamically stable with E$ _{hull} $ = 0eV, and Ti$ _{6} $In$ _{4} $ also has a relatively small E$ _{hull} $ = 0.121eV, which may be synthesized experimentally. Their structural information, E$ _{f} $ and E$ _{hull} $ are summarized in Table S2. 

\begin{table}[t]
	\renewcommand\arraystretch{1.15}
	\caption{Electronic DOS at the Fermi energy N(E$_F$) (states/(eV·f.u.)), EPC $\lambda$($\omega$ = $\infty$), logarithmically averaged phonon frequency $\omega_{log}$ and estimated Tc for Ti$_6$X$_4$.}
	{\centering
		\begin{tabular}{lp{1.3cm}<{\centering}p{3.0cm}<{\centering}p{0.7cm}<{\centering}p{1.8cm}<{\centering}p{1.1cm}<{\centering}p{1.1cm}}
			\hline
			\hline
			
		&Structures	& N(E$_F$) (states/(eV·f.u.))  & $\lambda$ &$\omega_{log}$ (K) &T$_c$ (K)\\
			\hline
									
			&Ti$_6$Bi$_4$   &5.707   &0.586    &177.3  	&3.8\\
			&Ti$_6$Sb$_4$  &5.153   &0.617    &187.8		&4.7\\
			&Ti$_6$Pb$_4$  &5.836   &0.642    &172.0		&4.8\\				
			&Ti$_6$In$_4$   &5.854   &0.604    &166.8  	&3.9\\
			&Ti$_6$Tl$_4$  &6.225   &0.666    &166.0		&5.1\\	

			\hline
			\hline
	\end{tabular}}
	\label{Table1}
\end{table}

\begin{table}[b]
	\renewcommand\arraystretch{1.15}
	\caption{Three independent components of SHC ($ \hbar $·(e·$ \Omega $·cm)$^{-1} $) at the Fermi energy for Ti$_6$X$_4$.}
	{\centering
		\begin{tabular}
			{lp{1.8cm}<{\centering}p{1.15cm}<{\centering}p{1.15cm}<{\centering}p{1.15cm}<{\centering}p{1.15cm}<{\centering}p{1.15cm}<{\centering}p{1.15cm}}
			\hline
			\hline
			
			&SHC $ \hbar $·(e·$ \Omega $·cm)$ ^{-1} $	&Ti$_6$Bi$_4$  &Ti$_6$Sb$_4$ &Ti$_6$Pb$_4$ &Ti$_6$In$_4$ &Ti$_6$Tl$_4$\\
			\hline
			&$ \sigma_{xy}^{z} $ &-160  &-225   &-420    &-274  	&-201\\
			&$ \sigma_{zx}^{y} $ &-333  &-629      &-230         &-66		&   80\\
			&$ \sigma_{yz}^{x} $ &-354  &-342      &-124         &34	  &-114\\						
			\hline
			\hline
	\end{tabular}}
	\label{Table2}
\end{table}

Similar to the analysis of Ti$ _{6} $Bi$ _{4} $, we also calculate the topological properties, electronic structures, superconducting properties and SHC of these Ti$ _{6} $X$ _{4} $ members. Ti$ _{6} $Sb$ _{4} $ has similar energy bands to that of Ti$ _{6} $Bi$ _{4} $. The DOS of Ti$ _{6} $Sb$ _{4} $ and Ti$ _{6} $Pb$ _{4} $ maintain the valley characteristic near E$ _{F} $ similar to that of Ti$ _{6} $Bi$ _{4} $, while the DOS of Ti$ _{6} $Tl$ _{4} $ and Ti$ _{6} $In$ _{4} $ show different behaviors. As plotted in FIGs. S3 - S6, abundant TSS are obtained in projected spectral functions for all members of Ti$ _{6} $X$ _{4} $, and the bands near Fermi level with a nonzero $\mathbb{Z}$$_2$ index indicate that they are topologically nontrivial, which show that these compounds are also $\mathbb{Z}$$_2$ topological metals. To study the superconductivity of the members of Ti$ _{6} $X$ _{4} $, we have calculated $\alpha$$^2$F($\omega$), $\omega_{log}$, EPC $\lambda$ and Tc at ambient pressure, as listed in TABLE.~\ref{Table1}. All other members of Ti$ _{6} $X$ _{4} $ have higher Tcs than Ti$ _{6} $Bi$ _{4} $, among which Ti$ _{6} $Tl$ _{4} $ has the highest Tc of 5.1K. Ti$_6$Sb$_4$ has a larger $\omega_{log}$ due to the higher vibrational frequency of lighter Sb atom, leading to a larger Tc. Ti$ _{6} $Pb$ _{4} $, Ti$ _{6} $In$ _{4} $ and Ti$ _{6} $Tl$ _{4} $ have higher Tcs mainly because they are equivalent to hole doping of Ti$ _{6} $Bi$ _{4} $, resulting in higher DOS and EPC $\lambda$. The energy bands of other members of Ti$ _{6} $X$ _{4} $ in FIGs. S4-7 also exhibit abundant gapped nodes and DNLs similar to Ti$ _{6} $Bi$ _{4} $, which contribute the SHC around 34-629 $ \hbar $·(e·$ \Omega $·cm)$ ^{-1} $ at E$ _{F} $ as listed in TABLE.~\ref{Table2}. These values are comparable to some reported compounds with high SHC, such as V$ _{6} $Sb$ _{4} $ [204-537 $ \hbar $·(e·$ \Omega $·cm)$ ^{-1} $] \cite{RN433}, Bi$ _{1-x} $Sb$ _{x}$ [474 $ \hbar $·(e·$ \Omega $·cm)$ ^{-1} $] \cite{RN498} and (Mo/W)Te$ _{2} $ [18-361 $ \hbar $·(e·$ \Omega $·cm)$ ^{-1} $] \cite{RN500}. The excellent stability, high SHC and the combination of high Tc and nontrivial topology make kagome family Ti$ _{6} $X$ _{4} $ worth exploring experimentally.

\section{IV. DISCUSSIONS}
We notice that a cousin materials of AV$ _{3} $Sb$ _{5}$ $ - $ V$ _{6} $Sb$ _{4} $, which share the same protype structure with Ti$ _{6} $X$ _{4} $, has also been synthesized recently \cite{RN350,RN429}. To explore the reason why in experiment V$ _{6} $Sb$ _{4} $ does not show signal of superconducting transition under pressure of 0$ - $80GPa \cite{RN350}, we analyze the magnetic properties of V$ _{6} $Sb$ _{4} $ using the same calculation method for Ti$ _{6} $X$ _{4} $. We find that V$ _{6} $Sb$ _{4} $ is a ferromagnet with a small magnetic moment of 0.35$ \mu_B$/f.u., which may explain the disappearance of superconductivity. Consistent with our calculations, the experimental magnetic susceptibility measurements showed that a small effective magnetic moment can indeed be measured despite the presence of impurities in V$ _{6} $Sb$ _{4} $ \cite{RN429}. In contrast, the nonmagnetic kagome family members Ti$ _{6} $X$ _{4} $ show high Tc, which deserves further experimental study.

To study the superconductivity in Ti$ _{6} $Bi$ _{4} $ as a function of pressure, we calculate its Tc at high pressures and our results show that the pressure strongly suppresses its Tc (see TABLE S1). On the other hand, it is expected that the Tc of Ti$ _{6} $Bi$ _{4} $ can be enhanced by doping. It is noteworthy that the electronic DOS of Ti$ _{6} $Bi$ _{4} $ is located in a valley at the E$ _{F} $ in FIG.~\ref{fig2}(a). By either electron or hole doping, the DOS can be greatly increased, and Tc may be improved correspondingly. So, the substitutional doping of Bi with elements of adjacent IIIA, IVA and VIA groups is a promising carrier dopant, which slightly changes the kagome nets of Ti atoms and the band structure near the Fermi level. These predictions can be checked by further experiments.

The coexistence of the superconducting ground state and clear TSS near E$ _{F} $ in Ti$ _{6} $Bi$ _{4} $ is similar to the behaviors of CsV$ _{3} $Sb$ _{5}$. Since the possible Majorana bound state is discussed in the experiment of CsV$ _{3} $Sb$ _{5} $ \cite{RN1}, the Majorana bound state is also expected in Ti$ _{6} $Bi$ _{4} $ due to the proximity effect. On the other hand, we notice that the Fermi surface of Ti$ _{6} $Bi$ _{4}$ encloses five TRIM of F, F$ _{1} $, F$ _{2} $, $ \Gamma $ and T as shown in FIGs.~\ref{fig2}(c, d), which shows a similar characteristic like odd-parity superconductors Sn$ _{1-x} $In$ _{x} $Te \cite{RN584} and Cu$ _{x} $Bi$ _{2} $Se$ _{3} $ \cite{RN441, RN532}. A concise theorem shows that an odd-parity superconductor with inversion symmetry is a TSC if its Fermi surface encloses an odd number of TRIM in the BZ \cite{RN441,RN538, RN539}. Although we only calculate superconductivity based on the traditional s-wave paired BCS theory, Ti$ _{6} $Bi$ _{4} $ is likely to have an odd-parity pairing potential beyond traditional s-wave pairing due to its strong SOC like Sn$ _{1-x} $In$ _{x} $Te \cite{RN584} and Cu$ _{x} $Bi$ _{2} $Se$ _{3} $ \cite{RN441,RN535}.  Under this presumption of odd-parity superconductor, although Tc of Ti$ _{6} $Bi$ _{4}$ may be changed, the characteristic of the Fermi surface indicates that Ti$ _{6} $Bi$ _{4} $ is a strong TSC. Therefore, Ti$ _{6} $Bi$ _{4}$ can serve as a promising platform for investigating Majorana zero-modes and TSC.

\section{V. CONCLUSIONS}

To summarize, we predict a promising kagome family $ - $ Ti$ _{6} $X$ _{4} $ (X = Bi, Sb, Pd, Tl, In) by DFT calculations. The thermodynamic and dynamic stability of these compounds is corroborated by the calculations of energy and phonon spectra. All members of Ti$ _{6} $X$ _{4} $ produce a superconducting transition with a Tc of 3.8$ - $5.1K and have a strong $\mathbb{Z}$$_2$ index with clear TSS near the Fermi level. The calculated spin texture of Ti$ _{6} $Bi$ _{4} $ shows TSS with spin helicity. Either the proximity-induced s-wave pairing on the surface or the possible odd-parity pairing with strong topological character show that Ti$ _{6} $Bi$ _{4} $ is a promising TSC. Based on the Kubo formula, the SHC of Ti$ _{6} $X$ _{4} $ is calculated to be about 34$ - $639 $ \hbar $·(e·$ \Omega $·cm)$ ^{-1} $. The large SHC is attributed to the large spin Berry curvature caused by the gapped nodes and DNLs. With high EPC superconductivity, excellent topological properties and large spin Hall effect, Ti$ _{6} $X$ _{4} $ deserve further experimental studies on their topological superconductivity and electronic transport properties.

\section* {DATA AVAILABILITY}
The data that support the findings of this study are available from the corresponding authors upon reasonable request.

	\section* {Acknowledgments}
This work is supported in part by the National Key R$ \& $D Program of China (Grant No. 2018YFA0305800), the Strategic Priority Research Program of the Chinese Academy of Sciences (Grants No. XDB28000000), the National Natural Science Foundation of China (Grant No.11834014), and High-magnetic field center of Chinese Academy of Sciences. B.G. is supported in part by the National Natural Science Foundation of China (Grant No. 12074378), the Chinese Academy of Sciences (Grants No. YSBR-030, No. Y929013EA2), the Strategic Priority Research Program of Chinese Academy of Sciences (Grant No. XDB33000000), and the Beijing Natural Science Foundation (Grant No. Z190011).

\section* {Author Contributions}	
G.S. designed and supervised the research. X.W.Y. performed theoretical calculation. All of the authors participated in analysing results. X.W.Y., J.Y.Y., B.G., and G.S. prepared the figures and the manuscript.

\section* {Competing interests}	
The authors declare that they have no competing financial or non-financial interests.
	

%

\end{document}